\documentclass[conference]{IEEEtran}
\IEEEoverridecommandlockouts
\usepackage{cite}
\usepackage{amsmath,amssymb,amsfonts}
\usepackage{algorithmic}
\usepackage{graphicx}
\usepackage{textcomp}
\usepackage{xcolor}
\usepackage{url}
\usepackage{array}
\graphicspath{ {./images/} }
\def\BibTeX{{\rm B\kern-.05em{\sc i\kern-.025em b}\kern-.08em
    T\kern-.1667em\lower.7ex\hbox{E}\kern-.125emX}}
\begin{document}

\title{From Channel Measurement to Training Data for PHY Layer AI Applications\\
\thanks{This work has been supported by the Federal Ministry of Education and Research of the Federal Republic of Germany as part of the FunKI project with grant number 16KIS1182.}
}
%

\author{%
    \IEEEauthorblockN{%
    	  Michael Zentarra,
      Julian Ahrens and
      Lia Ahrens
      }%
    \IEEEauthorblockA{%
      \begin{tabular}{c}%
        \textit{Intelligent Networks}\\%
        \textit{German Research Center for Artificial Intelligence}\\%
        Kaiserslautern, Germany%
      \end{tabular}%
      \qquad%
      \\%
      \{michael.zentarra,julian.ahrens,lia.ahrens\}@dfki.de}}

\maketitle

\begin{abstract}
Learning-based techniques such as artificial intelligence (AI) and machine learning (ML) play an increasingly important role in the development of future communication networks. The success of a learning algorithm depends on the quality and quantity of the available training data. In the physical layer (PHY), channel information data can be obtained either through measurement campaigns or through simulations based on predefined channel models. Performing measurements can be time consuming while only gaining information about one specific position or scenario. Simulated data, on the other hand, are more generalized and reflect in most cases not a real environment but instead, a statistical approximation based on a mathematical model. This paper presents a procedure for acquiring channel data by means of fast and flexible software defined radio (SDR) based channel measurements along with a method for a parameter extraction that provides configuration input to the simulator. The procedure from the measurement to the simulated channel data is demonstrated in two exemplary propagation scenarios. It is shown, that in both cases the simulated data is in good accordance to the measurements.
\end{abstract}


\section{Introduction}

In research and industry ongoing effort is devoted to further improvement of future wireless networks in order for these to fulfill the continuously growing requirements. Among various aspects being considered, AI is seen as one of the most potent technological enablers for 6G~\cite{Wei_6GSurvey}. AI and ML techniques are thereby expected to play a vital role in different parts of the network. One focus is the use in the physical layer. This is underlined by the work of 3GPP on a study item on AI/ML for new radio (NR) air interface in release 18.

The processes in the physical layer are traditionally viewed in a block structure separating the distinct tasks such as coding, modulation and mapping. While there are many well-established techniques that are optimized for their respective tasks and therefore provide near-optimal solutions, end-to-end optimization of communication systems is expected to allow further improvement~\cite{Wang_DLforPL}. The development of traditional algorithms is driven by theoretical work along with proper mathematical modelling of the tasks within the considered communication system, which can be especially difficult when joint optimization of multiple tasks shall be considered. In contrast, data-driven algorithms such as AI and ML, function in a data-adaptive manner, which allows for a more holistic approach.

Based on the above considerations, extensive training data of good quality are necessary for AI-based algorithms to provide benefits to the communication system. In general, channel data, which contains information about the state of the wireless propagation channel, can be obtained by means of measurement or simulation. While measurements can realistically depict the wireless channel, the associated hardware requirements as well as the effort are high. Furthermore, the amount of data is limited to the chosen measurement locations. Simulations, on the other hand, can generate data in any quantity. The quality of those data, however, may be lower, depending on the simulation method, because the simulations are in most cases based on generalized models that generate data according to stochastic distributions.

This paper provides a procedure where both channel measurement and simulation are combined. Therewith, some of the disadvantages of the measurements are mitigated by the use of SDRs, which enables a fast and flexible characterization of the channel. Subsequently, this measurement output is used to improve the accuracy of the simulation for a specific environment.

The structure of this paper is as follows: Section~\ref{sec: AIinPL} names AI applications in the PHY layer and further motivates the need for useful datasets. Section~\ref{ch: Data Generation} describes the different means to acquire such datasets. The proposed procedure is presented in section~\ref{sec: Meas and Sim}, together with the results recorded in two exemplary scenarios. Section~\ref{chap: conc} concludes the paper.

\section{AI Usage in the Physical Layer} \label{sec: AIinPL}


%
In recent years, a number of possible applications of ML methods in the PHY layer of communications systems have been proposed~\cite{OH17,Sim18}.
Therein, algorithms based on ML tools such as artificial neural networks replace the traditional signal processing blocks from which the communications system is built.
Compared to classical methods, the ML-based components usually require more computing power to be effective.
However, it is oftentimes still beneficial to employ ML methods, considering the significant performance improvement, particularly in cases where no theoretical solution to the considered problem is known, either because of inherent complexity of the problem itself or because of the extensive amount of unknown quantities involved, as with most real-world scenarios.
The applications under investigation range over AI-driven design of channel codes~\cite{JKA+19}, integrated transceiver design~\cite{AH18,FCD+18}, and channel state estimation and prediction~\cite{Ahrens_CNNFore,JS20}.
While some of these applications allow for the exclusive use of unsupervised learning in the sense that training does not require a dataset, for channel estimation and prediction, it is essential to have channel data that are both precise and diverse.
In fact, it is not reasonable to assume that superior performance can be achieved by a trained ML model if the data encountered during training can already be accurately captured by a much simpler statistical channel model.
Instead, the training data have to encompass all essential characteristics of the environment in which the system is likely to be placed and, at the same time, they need to be accurate enough to be used as ground truth during training.
This is particularly important when the ML components are based upon advanced artificial neural networks with high capacity such as the convolutional-type channel predictors for wide-band channels proposed in~\cite{Ahrens_CNNFore}.

\section{Training Data Generation} \label{ch: Data Generation}

\subsection{Data for Deep Learning} \label{sec: }

In the past, the introduction of modern ML methods has led to remarkable results in a variety of fields.
To facilitate the development and refinement of these algorithms and for comparative evaluation, many standard datasets have been collected, especially in fields where machine learning approaches have proven to be particularly beneficial.
E.g., MNIST, CIFAR10, and ImageNet for image processing, Penn Treebank and Hutter Prize for natural language processing, and CMUdict and Librispeech for speech recognition and production.
As the research on possible applications of ML to communications still has a rather short history, so far, no standard datasets have been established.
Such data may be obtained from two different kinds of sources: large-scale real-world measurement campaigns and advanced channel modelling and simulation.
In the following two sections both of these approaches are explored independently.
 
\subsection{Channel Measurements} \label{sec: Measurement}

The wireless channel can be arbitrarily complex due to large-scale fading effects such as shadowing and small-scale fading effects originating from reflected or scattered multipath components (MPCs) on objects as well as doppler shifts caused by transmitter (TX) and receiver (RX) movements. While the large-scale fading effects are relatively constant over a certain area and time, small-scale fading causes rapid fluctuations of signal strength even in small spatial or temporal areas. 

Obtaining channel data by the means of measurements provides the advantage of realism, only restricted by the accuracy and hardware impairments of the measurement setup. Furthermore, measurement data can contain information about several channel characteristics which can be used depending on the property of interest. However, to perform channel measurement campaigns, high-end hardware as well as a lot of time and effort are needed. Due to the changing character of wireless channels it is necessary to measure several measurement points in time and space to be able to characterize a measured environment. This leads to large amounts of data which are then difficult to distribute and make usable for the scientific community. Additionally, the measurements only capture the specific measured environment at the time of the measurements. 

When measuring only large-scale parameters like the signal strength, as it is for example done in~\cite{Zhou_DLSNR}, where the signal-to-noise ratio (SNR) is measured over a time period of several minutes in 10~ms timesteps in order to use it as training data for a deep-learning (DL) based SNR predictor, the requirements on the employed hardware are relatively low. However, in order to capture the small-scale parameters of the channel, the time resolution of the measurement setup has to be significantly higher, which increases the requirements on the employed hardware. Usual channel impulse responses (CIRs) in urban environments span over a time of 5-30~µs. A time resolution in the lower two-digit nanoseconds area or below is necessary to resolve the incoming MPCs. This is especially true for scenarios with small TX-RX separation distances. With the increasing focus on small cells, millimeter waves as well as indoor and industrial scenarios, all of which decrease the TX-RX distance compared to the urban or even rural scenarios in previous generations of mobile communication systems, the requirements on the measurement hardware continuously increases. Furthermore, the diversity of the considered scenarios, which differ significantly in terms of channel characteristics, also raises the effort for measurement campaigns to provide datasets for all scenarios. Besides the task of measurement, the additional overhead to openly distribute the obtained datasets, including data cleaning, descriptions of the measurement setup and environment as well as tutorials on how to use the data, leads to the situation that only few datasets are available, which can be used as training data for AI applications in the PHY layer.

One example of openly available measured channel datasets comes from the University of Stuttgart, who built a channel sounder called DICHASUS that measures massive MIMO channels in different environments such as indoor, outdoor or industrial~\cite{Euchner_DICHASUS}. They distribute some of their results together with tutorials and examples on a website to use for the scientific community.

Most measurement campaigns, however, focus on the characterization of a certain channel scenario in order to generalize to other locations and create channel models that describe the respective scenario as realistically as possible. One of the most influential models is the WINNER II channel model~\cite{WINNER2}. It is based on extensive channel measurements, which are described in~\cite{WINNER2}, chapters 2.4 and 3.2. 

\subsection{Channel Modelling and Simulation} \label{sec: ModandSim}

Channel models, such as the aforementioned WINNER II, are mathematical descriptions of the physical channel based on the measured channel parameters. These models can be used by simulators to simulate transmissions in specified scenarios.  This is the most common way to create training data for ML applications. However, as mentioned in section~\ref{sec: AIinPL}, the quality of the training data, and therefore the quality of the used channel model, influences the performance of the ML algorithm. Consequently, a proper understanding of how the channel models and simulators generate the data is of importance.
Channel models can generally be divided into two categories: stochastic and deterministic channel models~\cite{Wang_5GMeas}.

Deterministic channel models are mostly based on ray-tracing. Thereby, the propagation of rays from TX to RX is calculated based on the physical laws of reflection and diffraction. For this, the propagation environment has to be digitally rebuilt for the ray-tracer to be able to capture all significant path components of the channel. The complexity of this task scales with the complexity of the propagation environment. Therefore, the accuracy of the deterministic model is proportional to the effort put into rebuilding the environment. This leads to the same problem as with measurements: the data are only accurate for the considered environment.

Stochastic channel models tackle this issue by design: they are generalizations of certain scenarios, that provide information about an average channel in the specific scenario. The generation of the data is based on stochastic distributions, e.g., the positions of the scatterers are created randomly such that the characteristics of the results conform to a distribution over the predefined channel parameters. There are several subtypes of stochastic channel models, some of which are described in more detail in~\cite{Wang_5GMeas}. The most common channel models are geometry based stochastic channel models (GBSMs), many based on the aforementioned WINNER II model. In contrast to purely stochastic models, such as simple tapped delay line models, GBSMs model individual scatterers explicitly and distribute them stochastically. The channel is then simulated based on the generated geometric environment.

\section{Measurement-aided Simulation} \label{sec: Meas and Sim}

\begin{figure}
\centering
\includegraphics[scale=0.3]{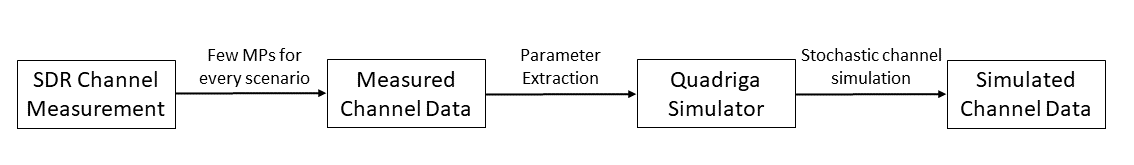}
\caption{System diagram of the procedure}
\label{pic:SD}
\end{figure}

This section describes the proposed procedure to acquire channel data out of measurement-aided simulations. A system diagram is depicted in Figure~\ref{pic:SD}. It starts with a SDR-based channel measurement step, which is described in more detail in section~\ref{sec: MeasSet and Det}. Out of the measured data some key parameters are extracted and used as a basis for channel simulation. These steps are described together with two exemplary scenarios in section~\ref{sec: ParamEx and Sim}. The simulator generates the desired channel data, which can be used as training data for AI applications.

\subsection{Measurement Setup and Details} \label{sec: MeasSet and Det}


The measurement step should be as fast and flexibly executable as possible since it is not intended as an exhaustive measurement campaign, but an assisting step towards the creation of channel data. Therefore, SDRs are used for the measurements. In combination with the open-source software GNU Radio a quick and easy deployment of a TX and RX are possible. Especially the use of the lightweight Ettus Research USRP B210, which can run on a USB connection to a laptop, allows for a flexible measurement setup in different environments~\cite{B210}.
Equipped with the TX and RX USRPs as well as their respective host PCs, single input single output (SISO) measurements are performed in the selected environments by transmitting and capturing a sounding signal at the frequencies 2.48~GHz and 5.75~GHz, which is sampled and saved with a sample rate of 25.6~MS/s. 

Due to the general purpose nature of SDRs, a tailored procedure is devised for their employment as a replacement of dedicated channel sounding hardware.
The proposed method comprises the following two components:
\begin{itemize}
  \item An iterative signal restoration routine counteracting the hardware specific artefacts encountered during data acquisition, where techniques from regression and time series analysis and anomaly detection come into use
  \item An encompassing time- and frequency-domain channel estimation procedure, with carefully chosen test signals based on a Zadoff-Chu sequence of prime length and post-processing stages, exploiting the dilation property of the Fourier transform so as to damp artefacts within the inferred channel impulse response
\end{itemize}
Empirical results demonstrate the efficacy of the proposed approach in both mitigating measurement artefacts and properly inferring the channel state information.
For more details on the method, the reader is referred to~\cite{Ahrens_SRCE}.

In each selected measurement environment several measurement points (MPs) with line-of-sight (LOS) and non LOS (NLOS) connections should be chosen to determine more significant channel parameters. Furthermore, for every MP three consecutive measurements were performed within a time interval of a few seconds to allow averaging. Following, two exemplary environments are considered: A location in urban Kaiserslautern as well as a campus-like environment at the DFKI building in Kaiserslautern. 

\begin{figure}
\centering
\includegraphics[scale=0.2]{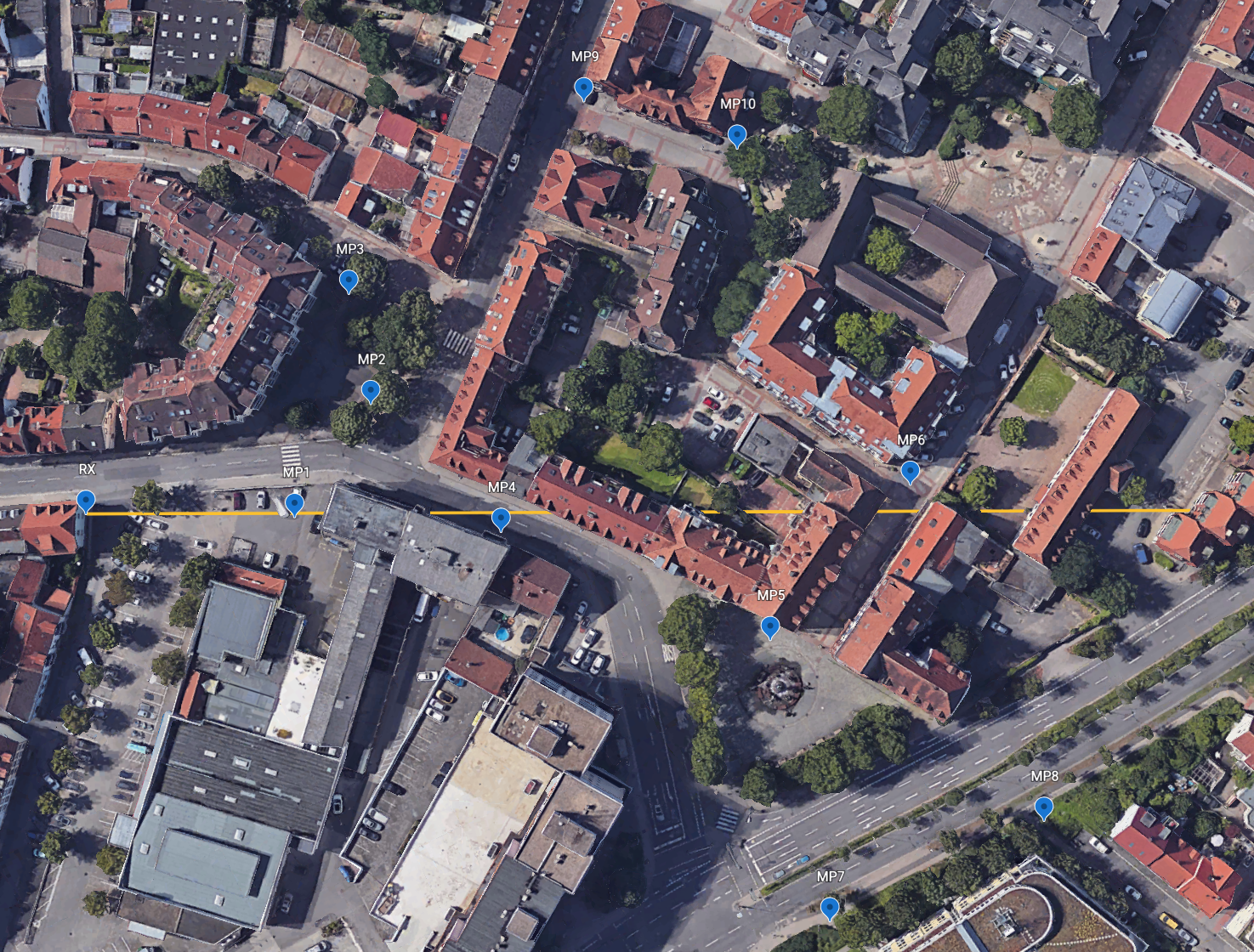}
\caption{Positions of the MPs in the urban measurements}
\label{pic:MP}
\end{figure}

The measurements in the urban environment included two LOS MPs for both frequencies as well as two NLOS MPs at 2.48~GHz and six NLOS MPs at 5.75~GHz. Figure~\ref{pic:MP} shows the exemplary MP distribution, whereby the MPs 1-4 were measured at both frequencies, the remaining only at 5.75~GHz. Since the attenuation at the MPs 6,7 and 10 was too strong, no meaningful results could be obtained there. All MPs range in distances between 50~m to 250~m. As can be seen on figure~\ref{pic:MP}, the environment is a typical urban scenario with multistory buildings and singular trees surrounding streets and a parking lot.

For the location at the DFKI building two LOS and two NLOS MPs were evaluated for each frequency. The location differs from the urban one in a way that there are only singular big buildings and a woodland part in the proximity.

\begin{figure}
\centering
\includegraphics[scale=0.09]{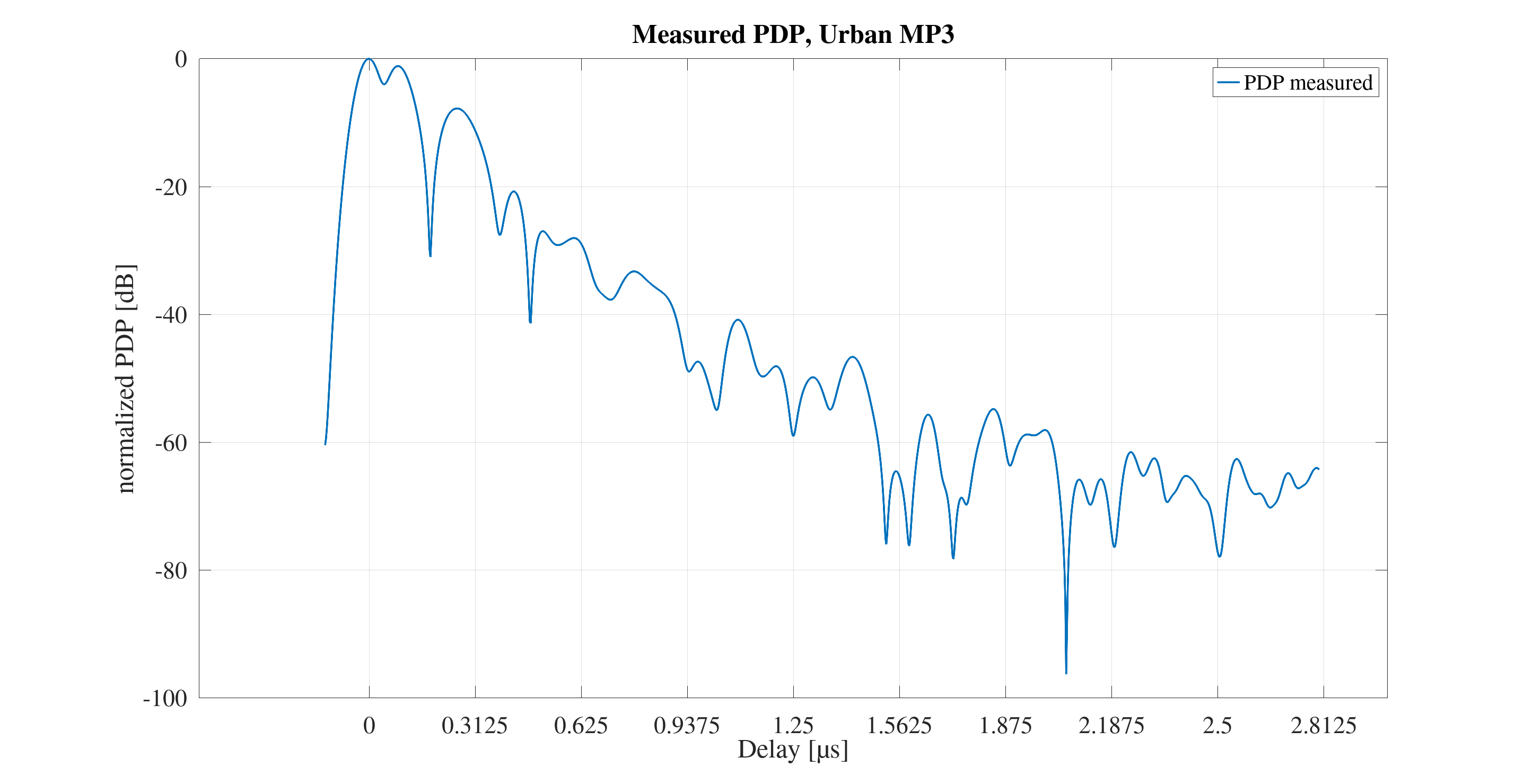}
\caption{Measured PDP at urban MP3}
\label{pic:MP3}
\end{figure}

One exemplary power delay profile (PDP) is depicted in Figure~\ref{pic:MP3}. One can see that the first peak is set to a delay of 0~µs. Subsequently, there are several peaks on a generally linear power decrease (on a logarithmic scale). It shows that singular MPCs can be resolved by the measurement. It has to be noted that the aforementioned channel estimation procedure includes an averaging step over several consecutive channel impulse responses, which results in a PDP~\cite{Rappaport}.

\subsection{Parameter Extraction \& Simulation} \label{sec: ParamEx and Sim}

For the simulation, the quasi deterministic channel generator (QuaDRiGa) developed at Fraunhofer HHI is used~\cite{Quadriga}. It is a geometry based stochastic channel model that allows simulation of realistic radio channels. Quadriga is compliant to the 3GPP-3D and NR channel models. It provides an open source MATLAB/GNU Octave framework, which includes several features to simulate from SISO to mMIMO channels such as dual mobility, spatial consistency and scenario transitions. It covers a frequency range from 450~MHz up to 100~GHz. Further details on the features as well as the simulation procedure and calculation can be found in~\cite{Quadriga_TR}.

One feature worth mentioning is the possibility to manually create semi-deterministic clusters. With this, one can add the most important scattering clusters according to their respective positions in the real environment. In this way the ray-tracing like nature of rebuilding parts of the real environment is adapted and combined with the stochastic approach for the remaining scatterers, while the effort is still kept low since the simulator will generate the remaining scatterers in a way to match the predefined channel parameters.

Quadriga uses configuration files according to the selected propagation scenario. These files include all the channel parameters for the respective scenario such as the delay spread, the K-factor, the number of scattering clusters, the angular spread and many more. For our purpose we can edit the channel parameters in the configuration file of the scenario, that fits the measured environment best, in order to improve the simulation results for our MP. For our scenarios the configurations were based on the configuratoin files "3GPP 38.901 UMi" for the urban case and "3GPP 38.901 UMa" for the campus scenario.

Out of our measurements it is possible to extract some key parameters of the channel, which can be put in the Quadriga configuration file afterwards. This includes the rms delay spread (DS), which is the second central moment of the PDP, defined as:
\begin{equation}\label{eq:rms_DS}
  \sigma_\tau = \sqrt{\overline{\tau^2}-(\overline{\tau})^2}
\end{equation}

with the mean excess delay
\begin{equation}\label{eq:MED}
  \overline{\tau} = \frac{\sum_k P(\tau_k)\tau_k}{\sum_k P(\tau_k)}
\end{equation}
and 
\begin{equation}\label{eq:DS2nd}
  \overline{\tau^2} = \frac{\sum_k P(\tau_k)\tau_k^2}{\sum_k P(\tau_k)}
\end{equation}

whereby $\tau$ is the delay after the first received signal part and $P(\tau)$ the respective power at a certain delay. Another value of interest is the ricean K-factor (KF), which is the ratio between the power of the strongest (usually LOS) path and the average power of the other, scattered paths. Both the DS and the KF are assumed to be log-normal distributed in Quadriga. Therefore, a median value as well as a standard deviation can be specified in the configuration file. Furthermore, the frequency, distance, height and elevation dependence on both the median and the standard deviation are possible parameters. In addition, the number of scattering clusters can be estimated by the amount of visible MPCs in the measured PDP. Because we executed SISO measurements, we cannot obtain information about the angular parameters.

The extracted channel parameters for the respective scenarios can be found in table~\ref{table: params}. The LOS and NLOS cases differ by significantly higher delay spreads as well as a higher number of visible scattering clusters. The differences between the urban and the campus scenario are minor, but generally all parameters are higher for the campus case.

\begin{table}
\caption{Channel parameters extracted from the measurements}
\begin{center}
\begin{tabular}{||c c c c||} 
 \hline
 Scenario & DS [ns] & KF [dB]& \# clusters\\ [0.5ex] 
 \hline\hline
 Urban LOS & 45 & 13 & 15 \\ 
 \hline
 Urban NLOS & 125 & x & 19 \\
 \hline
 Campus LOS & 50 & 21 & 17 \\
 \hline
 Campus NLOS & 175 & x & 22 \\
 \hline
\end{tabular}
\end{center}
\label{table: params}
\end{table}

\begin{figure}
\centering
\includegraphics[scale=0.09]{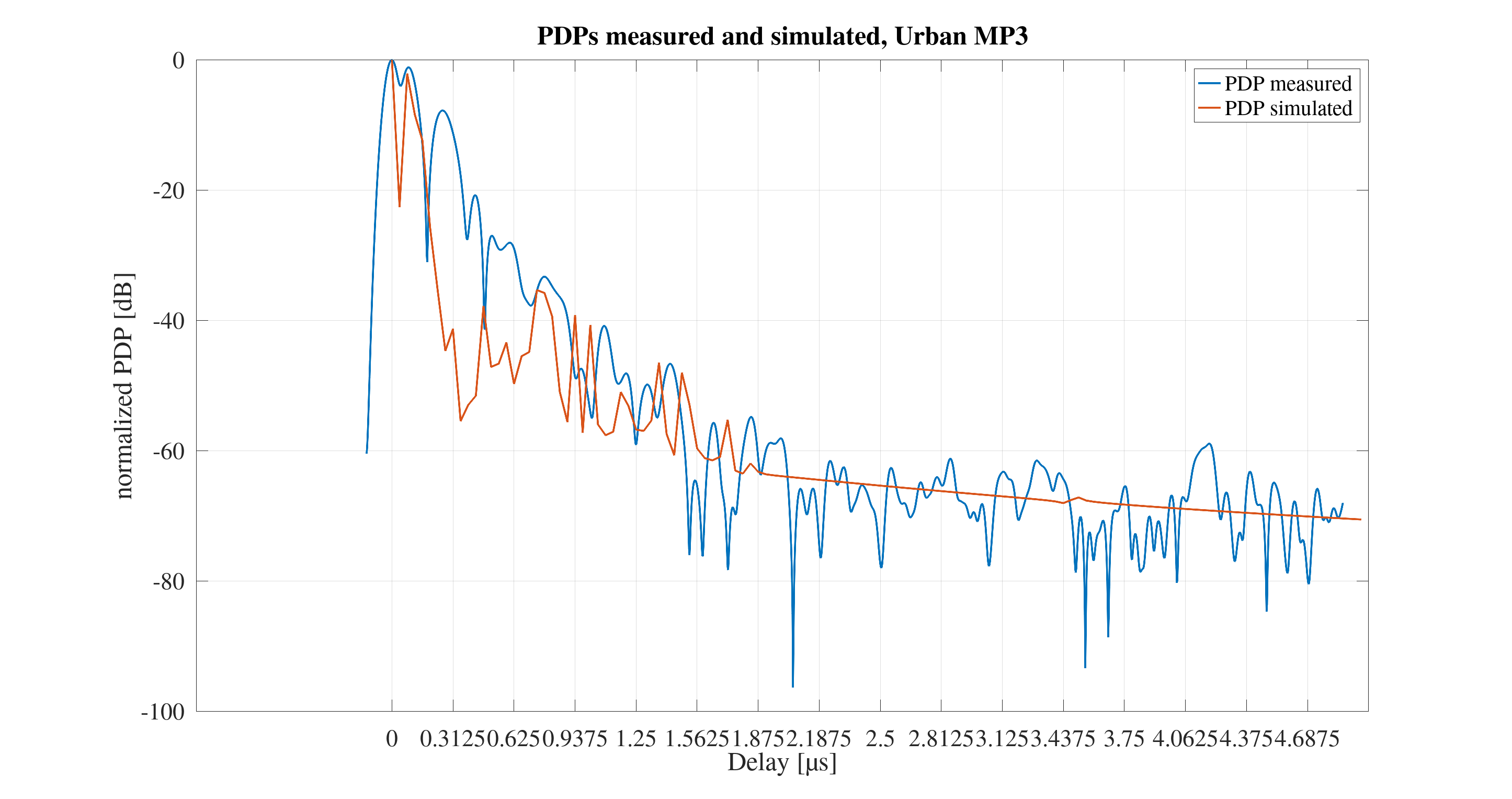}
\caption{Comparison of the measured and simulated PDP, urban MP3}
\label{pic:MP3_comp}
\end{figure}

\begin{figure}
\centering
\includegraphics[scale=0.09]{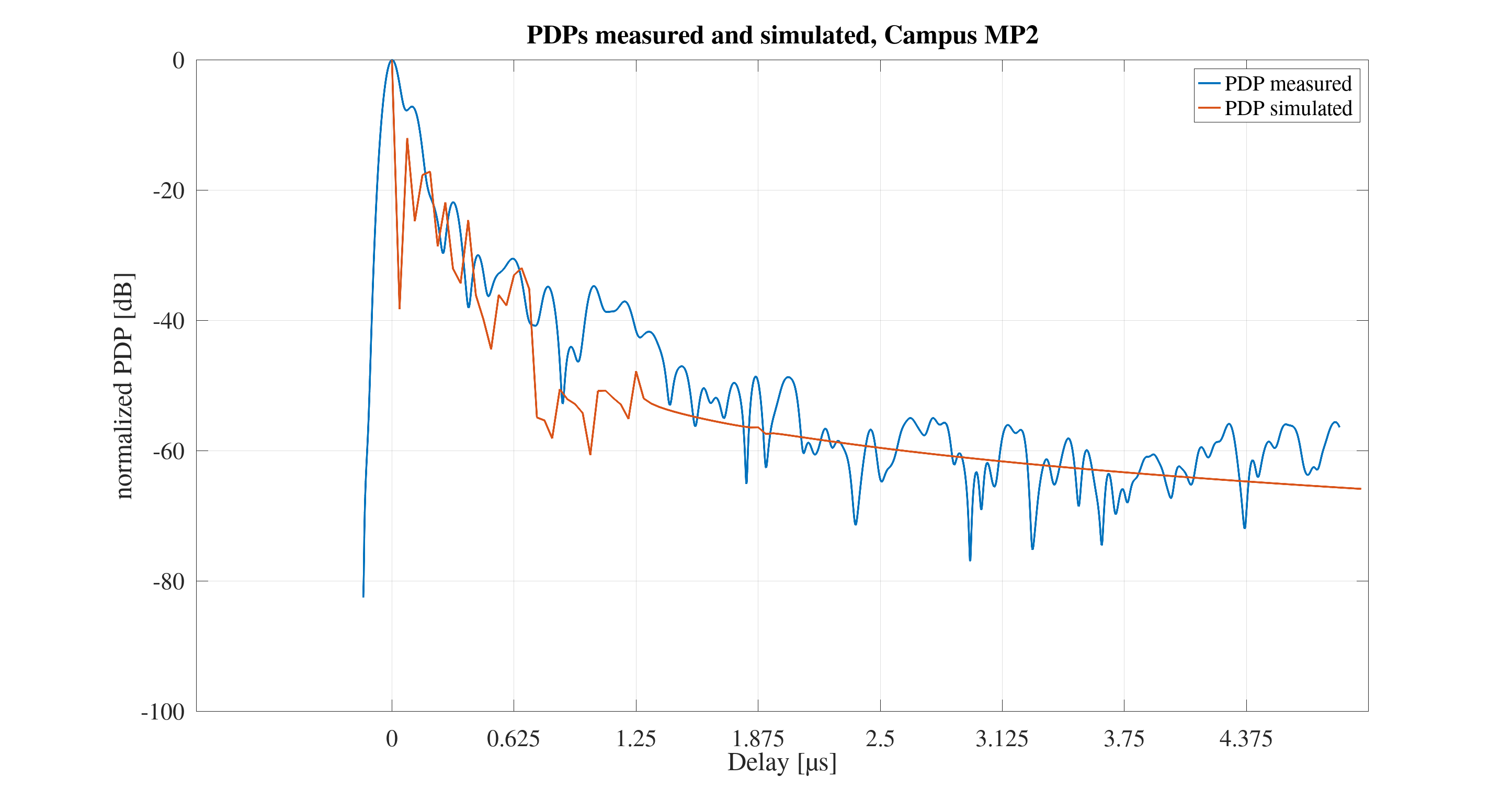}
\caption{Comparison of the measured and simulated PDP, campus MP2}
\label{pic:CampusMP2_comp}
\end{figure}

Two exemplary simulated PDPs are depicted in figure~\ref{pic:MP3_comp} and figure~\ref{pic:CampusMP2_comp}. In both cases, the blue graph represents the measured PDP of the respective channel.  They are normalized and the zero-delay is again set to the maximum value of the respective PDPs. The simulated PDPs, created by the described procedure with the parameters in table~\ref{table: params}, are shown in red. In the urban case in figure~\ref{pic:MP3_comp}, one can see that especially the first MPC of measurement and simulation fit well. Also in the delay time of 1 to 2~µs there is a similarity between both graphs.

The campus scenario in figure~\ref{pic:CampusMP2_comp} has a steeper early decrease in the simulated case. After a similar course around 0.5~µs the simulation did not capture the MPCs that were measured at around 1~µs delay.


It has to be noted, that the simulation of the PDPs is still a stochastic process depending on the instantiation of the simulator. Therefore, it cannot be expected that there is a perfect alignment between measurement and simulation. However, the procedure allows to bring in more accurate channel parameters as a simulation basis. This allows the assignment of a simulation to a certain location, which would not be possible in case of the use of the generalized channel models. Furthermore, based on the measured parameters the whole feature-set of the simulator can be used to create for example a mobility simulation with higher accuracy than with the default configuration.

\section{Conclusion} \label{chap: conc}

In this work, a procedure for acquiring training data for AI applications within a communication system is presented. The proposed method makes use of fast and flexible measurements to identify location-specific channel parameters which are passed on to simulators. In particular, the channel data are generated through the use of a combination of traditional ways, measurement and simulation. The measurement setup and execution is demonstrated in two exemplary scenarios. These measurements are used to configure the simulations based on the extracted channel parameters. Exemplary simulation results are presented, which show a good accordance between the measurements and the simulation.

\bstctlcite{IEEEexample:BSTcontrol}
\bibliographystyle{IEEEtran}
\bibliography{IEEEabrv,bib}
\end{document}